\documentclass[twocolumn,showpacs,preprintnumbers,amsmath,amssymb,prb]{revtex4}
\usepackage{bm}		
\usepackage{graphicx}	

\begin{document}


\title[The river model of black holes]{The river model of black holes}

\author{Andrew J. S. Hamilton}
\email{Andrew.Hamilton@colorado.edu}
\homepage{http://casa.colorado.edu/~ajsh/}
\author{Jason P. Lisle}
\affiliation{
JILA and Dept.\ Astrophysical \& Planetary Sciences,
Box 440, U. Colorado, Boulder CO 80309, USA
}

\def\slantfrac#1#2{\hbox{$\,^#1\!/_#2$}}	

\newcommand{\dd}{d}

\newcommand{\ff}{{\rm ff}}
\newcommand{\vff}{\beta}
\newcommand{\de}{d}
\newcommand{\ucoord}{\upsilon}
\newcommand{\utet}{u}

\newcommand{\eff}{{\rm eff}}

\newcommand{\avec}{\balpha}
\newcommand{\gvec}{\bm{g}}
\newcommand{\pvec}{\bm{p}}
\newcommand{\uvec}{\bm{u}}
\newcommand{\xvec}{\bm{x}}
\newcommand{\vffvec}{\bm{beta}}
\newcommand{\zhatvec}{\hat\bi{z}}
\newcommand{\gammavec}{\bm{\gamma}}
\newcommand{\partialvec}{{\bm \partial}}

\hyphenpenalty=3000

\newcommand{\blackholefishiesfig}{
    \begin{figure}
    \begin{center}
    \leavevmode
    \includegraphics[width=3in]{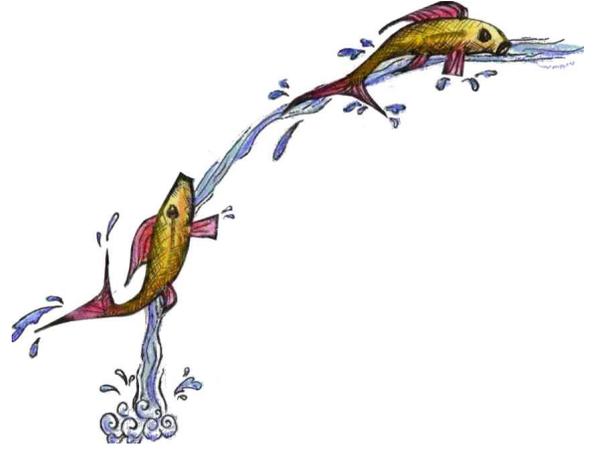}
    \end{center}
    \caption[1]{
    \label{blackholefishies}
(Color online)
The fish upstream can make way against the current,
but the fish downstream is swept to the bottom of the waterfall.
Figure~1 of \cite{Vachaspati04} presents a similar depiction.
    }
    \end{figure}
}

\newcommand{\tetradfig}{
    \begin{figure}
    \begin{center}
    \leavevmode
    \includegraphics[width=3in]{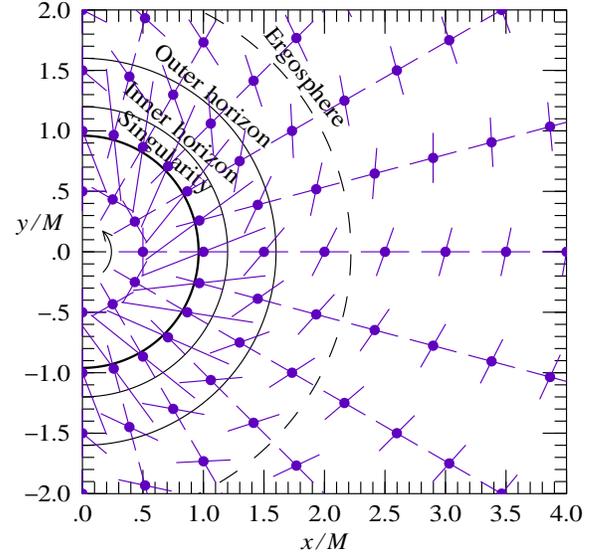}
    \end{center}
    \caption[1]{
    \label{tetrad}
(Color online)
Sets of horizontal radial and azimuthal tetrad axes
$\gammavec_\rightarrow$ and $\gammavec_\uparrow$,
equations~(\protect\ref{gammarperp}),
in the equatorial $x$-$y$ plane ($\theta = \pi/2$) of
an uncharged (Kerr) black hole with angular momentum per unit mass $a = 0.96$,
plotted in Doran-Cartesian coordinates.
The azimuthal axis at each point is tilted radially,
reflecting the fact that the spatial metric is sheared.
    }
    \end{figure}
}

\newcommand{\kerrfig}{
    \begin{figure}
    \begin{center}
    \leavevmode
    \includegraphics[width=3in]{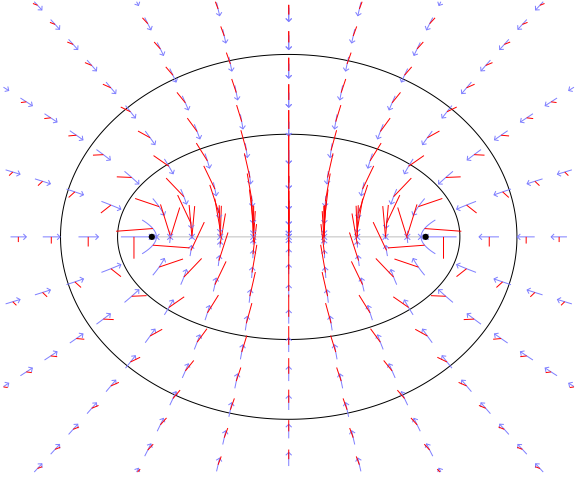}
    \end{center}
    \caption[1]{
    \label{kerr}
(Color online)
The velocity and twist fields for
an uncharged (Kerr) black hole with angular momentum per unit mass $a = 0.96$.
The arrowed lines show the magnitude and direction of the river velocity,
while the unarrowed lines emerging from the arrowed lines
show the magnitude and axis of the river twist.
The confocal ellipses show the outer and inner horizons,
and the large dots at the foci of the ellipses
indicate the ring singularity.
In the vacuum Kerr solution,
the river velocity goes to zero
at the horizontal disc bounded by the ring singularity,
then turns around and rebounds through a white hole into a new universe.
    }
    \end{figure}
}

\newcommand{\geodesicsfig}{
    \begin{figure}
    \begin{center}
    \leavevmode
    \includegraphics[width=3in]{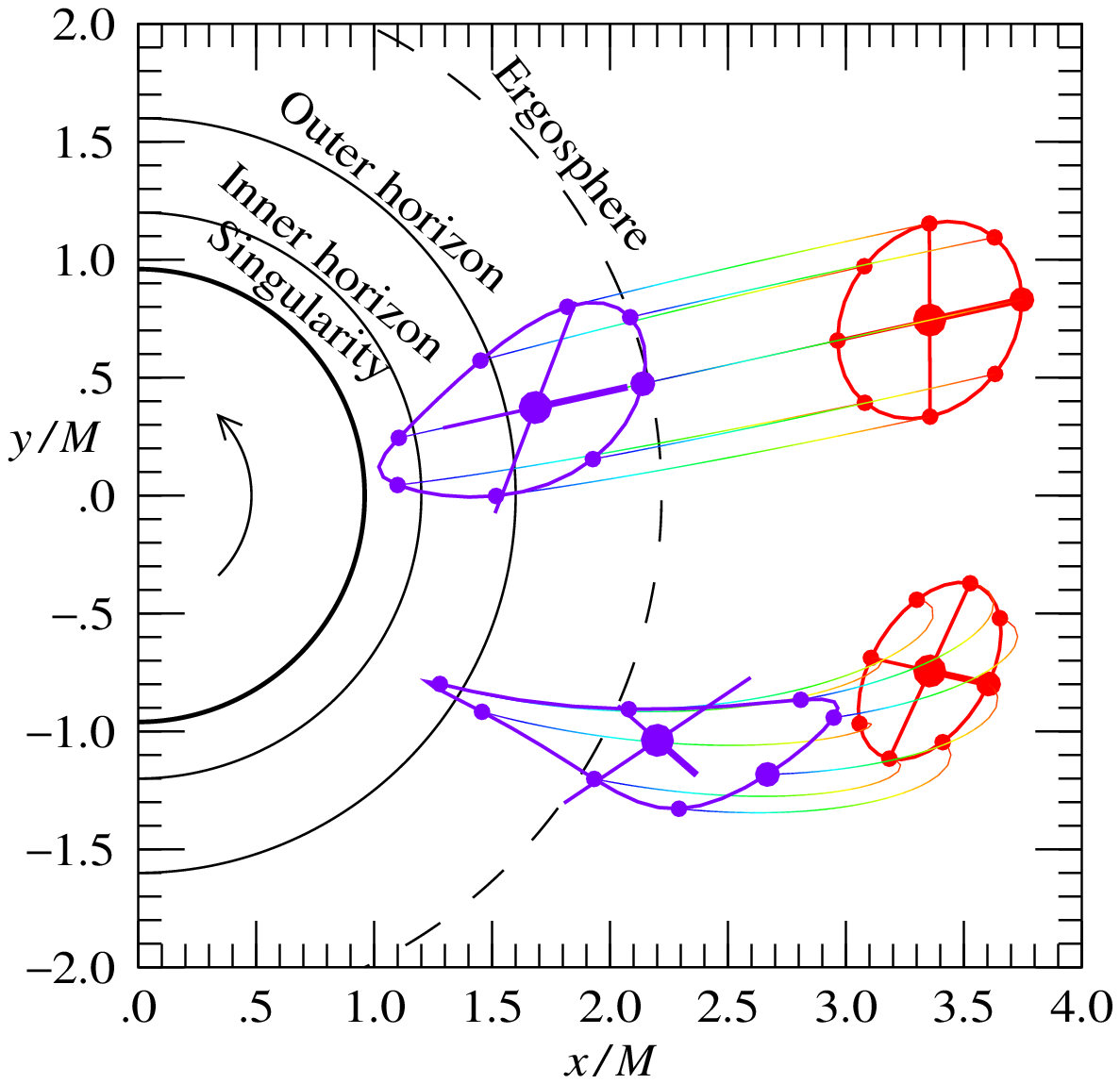}
    \end{center}
    \caption[1]{
    \label{geodesics}
(Color online)
Two sample sets of geodesics
in the equatorial $x$-$y$ plane ($\theta = \pi/2$) of
an uncharged (Kerr) black hole with angular momentum per unit mass $a = 0.96$,
plotted in Doran-Cartesian coordinates,
\S\protect\ref{dorancartesian}.
Each set of geodesics shows a central point (large dot)
and associated locally inertial axes
(crossed thick lines),
surrounded by a set of points that are
initially uniformly spaced around,
and initially at rest relative to, the central point,
in the locally inertial frame of the central point.
The central point and its attendants follow geodesics (thin lines)
into the black hole,
the ensemble becoming tidally distorted as it falls in.
In the upper set of geodesics,
the central point is comoving with the infalling river of space,
while
in the lower set of geodesics,
the central point is initially moving radially outward,
but soon turns around and falls in.
The lower ensemble illustrates how the locally inertial axes
attached to the central point twist as the ensemble falls in,
the twist acting so as to keep the frame comoving with
the geodesic motion of points in a small neighborhood
of the central point.
    }
    \end{figure}
}


\begin{abstract}
This paper presents an under-appreciated
way to conceptualize stationary black holes,
which we call the river model.
The river model is mathematically sound,
yet simple enough that the basic picture can be understood by non-experts.
In the river model,
space itself flows like a river
through a flat background,
while objects move through the river
according to the rules of special relativity.
In a spherical black hole,
the river of space falls into the black hole at the Newtonian escape velocity,
hitting the speed of light at the horizon.
Inside the horizon,
the river flows inward faster than light,
carrying everything with it.
We show that
the river model works also for rotating (Kerr-Newman) black holes,
though with a surprising twist.
As in the spherical case,
the river of space can be regarded as moving through a flat background.
However,
the river does not spiral inward, as one might have anticipated,
but rather falls inward with no azimuthal swirl at all.
Instead,
the river has at each point not only a velocity but also a rotation, or twist.
That is, the river has a Lorentz structure,
characterized by six numbers (velocity and rotation),
not just three (velocity).
As an object moves through the river,
it changes its velocity and rotation
in response to tidal changes in the velocity and twist
of the river along its path.
An explicit expression is given
for the river field, a six-component bivector field that encodes
the velocity and twist of the river at each point,
and that encapsulates all the properties of a stationary rotating black hole.
\end{abstract}

\pacs{04.20.-q}	

\maketitle

\section{Introduction}

As first pointed out in 1921 by
Allvar Gullstrand
\cite{Gullstrand}
and
Paul Painlev\'e
\cite{Painleve},
the Schwarzschild
\cite{AL04,Rothman02}
metric can be expressed in the form
\begin{equation}
\label{gullstrandmetric}
  \dd s^2
  =
  - \, \dd t_\ff^2 + (\dd r + \vff \, \dd t_\ff)^2
  + r^2 (\dd\theta^2 + \sin^2\!\theta \, \dd\phi^2)
\end{equation}
where
$\vff$ is the Newtonian escape velocity,
in units of the speed of light, at radius $r$
from a spherical object of mass $M$
\begin{equation}
  \vff = \left( {2 G M \over r} \right)^{1/2}
\end{equation}
and $t_\ff$ is
the proper time experienced by an object that
free falls radially inward from zero velocity at infinity.

Although Gullstrand's paper was published in 1922,
after Painlev\'e's,
it appears that Gullstrand's work has priority.
Gullstrand's paper was dated 25 May 1921,
whereas Painlev\'e's is a write up of a presentation to
the Acad\'emie des Sciences in Paris on 24 October 1921.
Moreover, Gullstrand seems to have had a better grasp of what he had discovered
than Painlev\'e,
for Gullstrand recognized that observables
such as the redshift of light from the Sun
are unaffected by the choice of coordinates in the Schwarzschild geometry,
whereas Painlev\'e, noting that the spatial metric was flat
at constant free-fall time, $\dd t_\ff = 0$,
concluded in his final sentence that,
as regards the redshift of light and such,
``c'est pure imagination de pr\'etendre
tirer du $\dd s^2$ des cons\'equences de cette nature''.

As shown in \S\ref{spherical},
the Gullstrand-Painlev\'e metric
provides a delightfully simple conceptual picture of
the Schwarzschild geometry:
it looks like ordinary flat space,
with the distinctive feature that space itself is flowing
radially inwards at the Newtonian escape velocity.
The place where the infall velocity hits the speed of light,
$\vff = 1$,
marks the horizon, the Schwarzschild radius.
Inside the horizon, the infall velocity exceeds the speed of light,
carrying everything with it.

\blackholefishiesfig

Picture space as flowing like a river into the Schwarzschild black hole.
Imagine light rays, photons, as fishes swimming fiercely in the current.
Outside the horizon,
photon-fishes swimming upstream can make way against the flow.
But inside the horizon, the space river is flowing inward so fast that it
beats all fishes, carrying them inevitably towards their ultimate fate, the
central singularity.

The river model of black holes offers a mental picture of black holes
that can be understood by non-experts
(at least in the spherical case)
without the benefit of mathematics.
It explains why light cannot escape from inside the horizon,
and why no star can come to rest within the horizon.
It explains how an extended object will be stretched radially
by the inward acceleration of the river,
and compressed transversely by the spherical convergence of the flow.
It explains why an object that falls through the horizon appears to an outsider
redshifted and frozen at the horizon:
as the object approaches the horizon,
light emitted by it takes an ever increasing time to forge against the
inrushing torrent of space and eventually to reach an outside observer.
The river model paints a picture that is radically different
from the Newtonian picture envisaged by
Michell (1784) \cite{Michell}
and Laplace (1799) \cite{Laplace}.

The picture of space falling like a river into a black hole
may seem discomfortingly concrete,
but the aetherial overtones are no more substantial
than in the familiar cosmological picture of space expanding
(see e.g.\ p.~237 of Greene 2004
\cite{Greene}).

As reviewed by Visser (1998, 2003) \cite{Visser98, Visser03}
and by Martel \& Poisson (2001) \cite{Martel},
the Gullstrand-Painlev\'e metric has been discovered and rediscovered repeatedly
\cite{Laschkarew26,Lemaitre33,Trautman66,Robertson68,EWCDST73,GH78,Gautreau95,Kraus94,Lake94,LDG98,NST99,Czerniawski02}.
Surprisingly,
the Gullstrand-Painlev\'e metric
is
widely neglected in texts on General Relativity.
An admirable exception is the
text ``Exploring Black Holes''
by Taylor \& Wheeler (2000) \cite{Taylor},
which devotes an entire section, Project B,
to the Gullstrand-Painlev\'e metric,
calling it the ``rain frame''
(the metric itself appears on page B-13).
Taylor \& Wheeler
attribute (page B-26)
the idea for the rain frame to
the book by Thorne, Price \& MacDonald (1986)
\cite{TPM86}, page 22 and elsewhere,
although the metric does not appear explicitly in the latter book.

It has been recognized for decades that
some aspects of general relativity can be conceptualized
in terms of flows.
In the
ADM (1962) formalism \cite{ADM62}
(see e.g.\ \cite{Lehner01} for a pedagogical review),
one considers fiducial observers -- FIDOs\cite{TPM86} --
whose worldlines are orthogonal to hypersurfaces of constant time.
The shift vector in the ADM formalism
is just the velocity of these FIDOs through the spatial coordinates.
Alcubierre (1994) \cite{Alcubierre94} constructed his
famous warp-drive metric by positing a superluminal
(faster-than-light) shift vector.

In a seminal, albeit initially unremarked paper,
Unruh (1981) \cite{Unruh81}
(see \cite{BLV05} for a comprehensive review)
pointed out that
the equations governing sound waves propagating in an inviscid,
barotropic (pressure is a specified function of density), irrotational fluid
are the same as those for a massless scalar field
propagating in a certain general relativistic metric.
Unruh showed that this implied that sound horizons
would emit Hawking radiation in much the same way as event horizons
in black holes,
and he proposed that Hawking radiation might be detected from
sonic black holes, or ``dumb holes'', in the laboratory.

As excellently reviewed by Barcel\'o et al.\ (2005) \cite{BLV05},
Unruh's paper in due course led to a now thriving industry
on ``analog gravity'',
in which fluid flows with prescribed velocity fields
simulate general relativistic spacetimes.
The primary aim of the work on analog gravity is to try to understand,
and perhaps in the not-too-distant future to probe experimentally,
quantum gravity through sonic analogs.



It is generally assumed that the fluid, or river, analogy
applies to a limited class of general relativistic spacetimes,
those in which the metric can be expressed
up to an overall factor (a ``conformal'' factor)
in terms of a shift vector
(the velocity of the river)
on an otherwise flat background space.
The 3-dimensional shift vector and the conformal factor
provide 4 degrees of freedom,
whereas at least 6 degrees of freedom are required to specify
an arbitrary spacetime
(the metric has 10 degrees of freedom,
of which 4 are removed by an arbitrary coordinate transformation).
As a corollary,
it has been thought that
any general relativistic geometry admitting a fluid analog
must necessarily be (up to a conformal factor) spatially flat at constant time
\cite{VW04,Natario04},
as indeed is the case in the
Gullstrand-Painlev\'e metric.

In particular,
it has been thought that no river model
for stationary rotating black holes
exists \cite{VW04},
since the Kerr-Newman geometry does not admit conformally flat slices
\cite{GP,Kroon}.

In the present paper we started from a somewhat different conceptual picture.
We noticed that fishes swimming in the Gullstrand-Painlev\'e river
moved according to the rules of special relativity,
being boosted by tidal differences in the river velocity
from place to place.
We wondered, might there be an analogous behavior for rotating black holes?
It came as a magical surprise, \S\ref{rotating}, that the answer is yes,
from this perspective
there is a river model of the Kerr-Newman geometry.
The rotating analog of the Gullstrand-Painlev\'e metric
proves to be
(as expected \cite{VW04})
the Doran (2000) \cite{Doran} form of the Kerr-Newman metric.
The new feature that emerges from the mathematics is that
the river of a rotating black hole is a fully 6-dimensional Lorentz river,
with a twist as well as a velocity.
Just as a velocity is a generator of a space-time rotation
(a Lorentz boost),
so also a twist is a generator of a space-space rotation
(an ordinary spatial rotation).
As a fish swims through the Doran river,
it is not only boosted but also rotated
by tidal differences in the river velocity and twist from place to place.

This novel point of view
leads to a different
notion of what is meant by the flat background space
through which the river flows and twists.
Mathematically, the essential feature of the river model
appears to be
equation~(\ref{gammaGammaomega}),
which states that the connection coefficients,
expressed in locally inertial frames
comoving with the infalling river of space,
should equal the ordinary (non-covariant) gradient of the river field.

The property
that the tetrad connection coefficients are equal to the ordinary gradient,
in Doran-Cartesian coordinates,
of the river field,
essentially defines what we mean by the background space
in the river model being flat.
This feature appears to be a special property of stationary black holes.
How this idea emerges from the mathematics is examined in
\S\ref{flatbkgd}, and revisited in \S\ref{flatbkgdre}.
We emphasize that the background being flat does not mean that
the metric is spatially flat,
although the latter is also true in the case of spherical black holes.
The notion that there is a sense in which stationary rotating black holes
admit a flat background coordinatization
might have application to numerical general relativity,
for example in setting up initial conditions containing rotating black holes,
where traditional conformal imaging and puncture methods that
assume a conformally flat 3-geometry are too restrictive to admit
Kerr black holes
\cite{Cook}.

Throughout this paper we adopt the sign conventions
and ordering of indices of
Misner, Thorne and Wheeler \cite{MTW}.

\section{Spherical black holes}
\label{spherical}

In this section we consider spherically symmetric black holes,
and we justify the assertion
that the Gullstrand-Painlev\'e metric, equation~(\ref{gullstrandmetric}),
can be interpreted as representing a river of space falling radially inward
at velocity $\vff$.
We demonstrate two features that are the essence of the river model
for spherical black holes:
first, that the river of space can be regarded as moving in Galilean fashion
through a flat Galilean background space
[eqs.~(\ref{gammaspher}) and (\ref{uspher})],
and second,
that as a freely-falling object moves through the flowing river of space,
its 4-velocity,
or more generally any 4-vector attached to the freely-falling object,
can be regarded as evolving by a series of infinitesimal Lorentz boosts
induced by the change in the velocity of the river from one place to the next
[eq.~(\ref{dpspher})].
Because the river moves in a Galilean fashion,
it can, and inside the horizon does,
move faster than light through the background.
However,
objects moving in the river move according to the rules of special relativity,
and so cannot move faster than light through the river.

\subsection{Mathematics of the river model}

In general, a spherically symmetric metric of the form
(units $c = G = 1$)
\begin{eqnarray}
\label{spher}
  \dd s^2
  &=&
  - \, \left[ 1 - 2 M(r) / r \right] \dd t^2
  + {\dd r^2 \over \left[ 1 - 2 M(r) / r \right]}
\nonumber
\\
  &&
  + \, r^2 (\dd\theta^2 + \sin^2\!\theta \, \dd\phi^2)
\end{eqnarray}
can be expressed in the Gullstrand-Painlev\'e form~(\ref{gullstrandmetric})
with infall velocity
\begin{equation}
  \vff(r) = \left[ {2 M(r) \over r} \right]^{1/2}
\end{equation}
the free-fall time $t_\ff$ being
\begin{equation}
  t_\ff = t - \int_r^\infty {\vff \over 1 - \vff^2} \, \dd r
  \ .
\end{equation}
The velocity $\vff$ is commonly called the shift in the
ADM formalism \cite{ADM62,Lehner01},
but in this paper we refer to $\vff$ as the river velocity.
The river velocity $\vff$ is positive for a black hole (infalling),
negative for a white hole (outfalling).
Horizons occur where the river
velocity $\vff$
equals the speed of light,
\begin{equation}
  \vff = \pm 1
  \ ,
\end{equation}
with $\vff = 1$ for black hole horizons,
and $\vff = -1$ for white hole horizons.
The Reissner-Nordstr\"om metric
for a spherically symmetric black hole
of mass $M$ and charge $Q$
takes the form~(\ref{spher}) with
mass $M(r)$ interior to $r$,
the so-called Misner-Sharp mass
(Misner \& Sharp 1964 \cite{MS}; Weinberg 1972, p~300 \cite{Weinberg}),
given by
\begin{equation}
\label{rn}
  M(r) = M - {Q^2 \over 2 r}
  \ .
\end{equation}
However, the river velocity $\vff$
can also be considered to be a more general function of radius $r$.
In \S\ref{reissnernordstrom}
we will return briefly to the Reissner-Nordstr\"om solution
to see what its river looks like.

To make the argument plainer,
rewrite the Gullstrand-Painlev\'e metric~(\ref{gullstrandmetric})
in Cartesian coordinates
$x^\mu = (x^0,x^1,x^2,x^3)$
=
$(t_\ff,x,y,z)$
instead of spherical coordinates:
\begin{equation}
  \dd s^2 =
  \eta_{\mu\nu}
  (\dd x^\mu - \vff^\mu \dd t_\ff) (\dd x^\nu - \vff^\nu \dd t_\ff)
\end{equation}
where $\eta_{\mu\nu}$ is the Minkowski metric,
and
\begin{equation}
  \vff^\mu
  =
  \vff \left(
    0 \, , \, - {x \over r} \, , \, - {y \over r} \, , \, - {z \over r}
  \right)
\end{equation}
are the components of the radial river velocity.

Let $\gvec_\mu$ denote the basis of tangent vectors
in the Gullstrand-Painlev\'e-Cartesian coordinate system $x^\mu$.
By definition,
the scalar products of the tangent vectors constitute the metric $g_{\mu\nu}$
\begin{equation}
\label{gg}
  \gvec_\mu \cdot \gvec_\nu = g_{\mu\nu}
  \ .
\end{equation}
Let $\ucoord^{\mu} \equiv \dd x^{\mu} / \dd \tau$
denote the 4-velocity of a particle falling freely (not necessarily radially)
in the geometry,
$\tau$ being the proper time experienced by the particle.
%
In particular,
observers who free-fall radially from zero velocity at infinity
have 4-velocity
\begin{equation}
  \ucoord_\ff^\mu = (1, \vff^1, \vff^2, \vff^3)
  \ .
\end{equation}
Such observers are comoving with the inflowing river of space.
Let
$\gammavec_m$,
and associated local coordinates $\xi^m = (\xi^0,\xi^1,\xi^2,\xi^3)$,
denote a system of locally inertial orthonormal frames, tetrads,
attached to observers who free-fall radially from zero velocity at infinity.
Here and throughout this paper we use latin indices
to signify tetrad frames,
reserving greek indices for curved spacetime frames.
Orthonormal means that
the scalar products of the tetrad basis at each point of spacetime
form the Minkowski metric
\begin{equation}
\label{gammaeta}
  \gammavec_m \cdot \gammavec_n = \eta_{mn}
  \ .
\end{equation}
That the tetrad frames move with the radially free-falling observers
without precessing requires that the vectors $\gammavec_m$
be `parallel-transported'
along the worldlines of these observers,
that is,
\begin{equation}
\label{detadtau}
  \ucoord_\ff^\mu \, {\partial \gammavec_m \over \partial x^\mu} = 0
  \ .
\end{equation}
Assume without loss of generality
that the tetrad frames $\gammavec_m$
are aligned with the Gullstrand-Painlev\'e-Cartesian frame at infinity.
Then the tetrad frames $\gammavec_m$
are related to
the Gullstrand-Painlev\'e basis $\gvec_\mu$ at each point by
\begin{equation}
\label{gammaspher}
  \begin{array}{ll}
  \gammavec_0 = \gvec_0 + \vff^i \gvec_i & \\
  \gammavec_i = \gvec_i & (i = 1,2,3)
  \end{array}
\end{equation}
which is most easily deduced from
equations~(\ref{gg}) and (\ref{gammaeta}) and considerations of symmetry,
the result being confirmed by checking that equation~(\ref{detadtau})
is then true.
Remarkably,
the relations~(\ref{gammaspher})
are those of a Galilean transformation,
which shifts the time axis by velocity $\vff$ along the direction of motion,
but leaves unchanged
both the time component of the time axis
and all of the spatial axes.

The 4-velocity $\utet^m$ of a freely-falling particle with respect to the
the tetrad frame $\gammavec_m$ at the position of the particle follows from
$\utet^m \gammavec_m = \ucoord^\mu \gvec_\mu$,
which implies
\begin{eqnarray}
\label{uspher}
  \utet^0 &=& \ucoord^0
\nonumber
\\
  \utet^i &=& \ucoord^i - \vff^i \ucoord^0
  \qquad (i = 1,2,3) \ .
\end{eqnarray}
Physically,
the 4-velocity $\utet^m$ is the 4-velocity of the particle relative
to the inflowing river of space.
For example,
the spatial components $\utet^i$ of the 4-velocity
are zero if the particle is becalmed in the river.
Again, the relations~(\ref{uspher})
resemble those of a Galilean transformation,
which shift only the spatial components of the vector,
while leaving the time component unchanged.
The only non-Galilean (relativistic) feature of equations~(\ref{uspher})
is that the 4-velocities $\utet^m$ and $\ucoord^\mu$
are derivatives with respect to proper time.
But proper time is a property of the objects moving in the river,
not of the river itself.
Objects moving in the river move through it
according to the rules of special relativity.
The river itself flows in Galilean fashion through a flat Galilean background.

We have demonstrated, equations~(\ref{gammaspher}) and (\ref{uspher}),
the first of the two claimed features of the river model for
spherical black holes,
that the river of space moves in Galilean fashion
through a flat Galilean background.
We proceed to demonstrate the second feature of the river model,
that objects moving in the river of space
move according to the rules of special relativity,
being Lorentz boosted by tidal differences in the river velocity
from place to place.

The tetrad frames have been constructed so that
an observer who free-falls from zero velocity at infinity
finds their own frame aligned at all times with the tetrad frame.
But in general another observer who free-falls along a different
geodesic will find their own locally inertial frame becoming misaligned
with the tetrad frame.
This misalignment is determined mathematically by the equations of motion
of objects, 4-vectors, expressed with respect to the tetrad frame.
Let $\pvec = p^m \gammavec_m = p^\mu \gvec_\mu$ be a 4-vector.
Its components $p^m$ in the tetrad frame are related to those $p^\mu$
in the coordinate frame by
$p^m = \delta_\mu^m p^\mu - \vff^m p^0$.
For a 4-vector in free-fall,
the equations of motion for the components $p^m$ in the tetrad frame are
(see \S\ref{eqmottetrad})
\begin{equation}
\label{dpdtauspher}
  {\dd p^k \over \dd \tau} + \Gamma^k_{mn} \utet^n p^m = 0
\end{equation}
where $\Gamma^k_{mn}$
are the tetrad frame connection coefficients,
the tetrad frame analog of the coordinate frame Christoffel symbols.
In the present case of spherically symmetric black holes,
the non-zero tetrad frame connection coefficients prove to be
given by the spatial gradient of the river velocity
(see \S\ref{motion})
\begin{equation}
\label{Gammaspher}
  \Gamma^0_{ij}
  =
  \Gamma^i_{0j}
  =
  {\partial \vff^i \over \partial x^j}
  \quad
  \qquad (i,j = 1,2,3)
  \ .
\end{equation}
From equations~(\ref{dpdtauspher}) and (\ref{Gammaspher}) it follows that
\begin{eqnarray}
  {\dd p^0 \over \dd \tau} &=&
  - {\partial \vff_i \over \partial x^j}
  \utet^j p^i
\nonumber
\\
\label{dpspher}
  {\dd p^i \over \dd \tau} &=&
  - {\partial \vff^i \over \partial x^j}
  \utet^j p^0
  \qquad (i = 1,2,3)
  \ .
\end{eqnarray}
The summations over paired indices in equations~(\ref{dpspher})
are formally over all four indices $0,1,2,3$,
but in practice reduce to sums over just the three spatial indices $1,2,3$
since, first,
the infall velocity has zero time component,
$\vff^0 = 0$,
as we have defined it,
and second,
the infall velocity has zero time derivative,
$\partial \vff^\mu / \partial x^0 = 0$.

In the context of the river model,
the equations of motion~(\ref{dpspher})
have the following interpretation.
In an interval $\delta \tau$ of proper time,
a particle moves a distance
$\delta x^i = \ucoord^i \delta \tau$
in the background Gullstrand-Painlev\'e-Cartesian coordinates,
and a proper distance
$\delta \xi^i = \utet^i \delta \tau = \delta x^i - \vff^i \delta t_\ff$
relative to the infalling river of space.
The proper distance $\delta \xi^i$ equals the distance $\delta x^i$
minus the distance $\vff^i \delta t_\ff$ moved by the river.
In Gullstrand-Painlev\'e-Cartesian coordinates,
the velocity $\vff^i$ of the infalling river at the new position differs
from the velocity at the old position by
$\delta x^j \, \partial \vff^i / \partial x^j$.
However,
in the river model,
a particle moving in the river sees not the full change in river velocity
relative to the background coordinates,
but only the tidal change
\begin{equation}
\label{dv}
  \delta \vff^i = {\partial \vff^i \over \partial x^j} \delta \xi^j
\end{equation}
in the river velocity relative to the infalling locally inertial river frame.
For example,
if the particle is comoving with the inflowing river,
so that $\delta \xi^i = 0$,
then the particle sees no change at all in the river velocity as time goes by,
$\delta \vff^i = 0$.
The infinitesimal tidal change $\delta \vff^i$ in the river velocity
induces a Lorentz boost in the 4-vector $p^m$
\begin{eqnarray}
\label{dp}
  p^0 &\rightarrow& p^0 - \delta \vff_i \, p^i
\nonumber
\\
  p^i &\rightarrow& p^i - \delta \vff^i \, p^0
  \qquad (i = 1,2,3)
  \ .
\end{eqnarray}
Equations~(\ref{dv}) and (\ref{dp})
reproduce the equations of motion~(\ref{dpspher}).



We have thus demonstrated the second of the claimed features
of the river model for spherical black holes,
that as a particle moves through the river of space,
its 4-velocity, or more generally any 4-vector attached to it,
is Lorentz boosted by tidal changes in the river velocity along its path.

\subsection{Reissner-Nordstr\"om metric}
\label{reissnernordstrom}

We conclude this section by commenting on
the Reissner-Nordstr\"om (RN) metric
for a spherical black hole of mass $M$ and charge $Q$.
In this case
the mass $M(r) = M - Q^2 / (2 r)$ interior to $r$, equation~(\ref{rn}),
can be interpreted as the mass $M$ at infinity, less the mass
$\int_r^\infty (E^2 / 8\pi) \, 4 \pi r^2 \dd r = Q^2/(2 r)$
contained in the electric field
$E = Q / r^2$
outside $r$.
The RN geometry
exhibits both outer and inner horizons $r_+$ and $r_-$
\begin{equation}
  r_{\pm} = M \pm (M^2 - Q^2)^{1/2}
  \ .
\end{equation}
The inflow velocity $\vff$
hits the speed of light at the outer horizon $r_+$,
reaches a maximum velocity
between the outer and inner horizons,
slows back down to the speed of light at the inner horizon $r_-$,
and slows all the way to zero velocity at the turnaround radius
\begin{equation}
  r_0 = {Q^2 \over 2 M}
  \ .
\end{equation}
At this point the flow of space turns around,
accelerates back outward through another inner horizon,
the so-called Cauchy horizon,
into a white hole,
and bursts through the outer horizon of the white hole into a new universe.

Sadly, the RN solution is not realistic,
and its promise of passages to other universes is moot.
The RN solution describes the geometry of empty space
surrounding a classical point charge of infinitesimal size,
and as such consists of
a gravitationally repulsive singularity of infinite negative mass
sheathed in an electric field containing an infinite positive mass
balanced so as to yield a finite mass $M$ at infinity.
See
\cite{HS04}
for an entry to the literature on the intriguing subject of what
really happens inside charged black holes.

The infall velocity $\vff$ is imaginary inside the turnaround radius $r_0$
of the Reissner-Nordstr\"om geometry,
the interior mass $M(r)$ being negative inside this radius.
This might be considered a defect of the river model,
but it might also be considered an asset,
signalling the presence of
unphysical negative mass.
Whatever the case,
the mathematical formalism remains valid
even where the river velocity $\vff$ is imaginary.

\section{Rotating black holes}
\label{rotating}

Does the river model work also for stationary rotating black holes?
As will be shown in this section,
the answer is yes:
there is a river of space, and it moves through a flat background,
and fishes move through the river special relativistically
as though they were being carried with it.
But the river has a surprising twist.
One might have anticipated that the river would spiral into the black hole
like a whirlpool, but that is not the case.
Rather, the river velocity has no azimuthal component at all.
Instead of a spiral,
the river possesses,
besides a velocity at each point,
a rotation, or twist, at each point.
The river is characterized not by three numbers, a velocity vector,
but by six numbers, a velocity vector and a twist vector.
As a fish swims through the river,
it is Lorentz boosted by gradients in the velocity of the river,
and rotated spatially by gradients in the twist of the river.


A key result of this section is the expression~(\ref{omega})
for the river field $\omega_{km}$.
This is a 6-component bivector \cite{MTW,GLD} field,
antisymmetric in its indices $km$,
whose electric part specifies the river velocity,
and whose magnetic part specifies the river twist.
The river field $\omega_{km}$
encapsulates all the properties of a stationary rotating black hole.


How can a river move and twist without spiralling?
The answer to this conundrum is that,
unlike the Gullstrand-Painlev\'e case,
the spatial metric is not flat, but sheared
(Fig.~\ref{tetrad}).
One can regard the twist in the river
as inducing the shear in the spatial metric;
or equally well one can regard the shear in the spatial metric
as requiring a twist in the river.
Whatever the case,
the twist and the shear act together
in just such a way as to ensure that locally inertial frames
moving through the infalling river
comove with the geodesic motion of points at rest
in a small neighborhood of the frame
(Fig.~\ref{geodesics}).

Recall from special relativity
that Lorentz transformations are generated
by a combination of changes in velocity,
or Lorentz boosts,
and spatial rotations.
Lorentz boosts are rotations in a plane defined by a space axis and a time axis,
while spatial rotations are rotations in a plane defined by two spatial axes.
Gradients in the velocity of the river
make the metric non-flat with respect to the time components,
while leaving the spatial metric at constant time flat.
Gradients in the rotation, or twist, of the river
make the metric non-flat with respect to the spatial components,
while leaving the time part of the metric flat,
that is, the metric becomes $- \, \dd t^2 + g_{ij} \dd x^i \dd x^j$
where $g_{ij}$ is a purely spatial metric.
We see that the reason that the Gullstrand-Painlev\'e metric
for spherical black holes
is flat along hypersurfaces of constant free-fall time
is attributable to the fact that the river has no twist component.
However, the Gullstrand-Painlev\'e river does have a velocity component,
so the Gullstrand-Painlev\'e metric is not flat in the time direction.
For rotating black holes,
the river has both velocity and twist components,
and the metric is flat neither in time nor in space.

\subsection{Doran metric}

Proceed to the mathematics.
Doran (2000)
\cite{Doran}
has pointed out that the
Kerr-Newman metric for a rotating black hole
of angular momentum $a$ per unit mass
(for positive $a$,
the black hole rotates right-handedly about its axis)
can be cast
in oblate spheroidal coordinates
$(t_\ff,r,\theta,\phi_\ff)$
in the form
\begin{eqnarray}
\label{doranmetric}
  \dd s^2
  &=&
  - \, \dd t_\ff^2
  + \left[ {\rho \dd r \over R}
  + {\vff R \over \rho} (\dd t_\ff - a \sin^2\!\theta \, \dd\phi_\ff) \right]^2
\nonumber
\\
  &&
  + \, \rho^2 \dd\theta^2 + R^2 \sin^2\!\theta \, \dd\phi_\ff^2
\end{eqnarray}
where $\vff(r)$ is the river velocity,
\begin{equation}
  R \equiv (r^2 + a^2)^{1/2}
  \ , \quad
  \rho \equiv (r^2 + a^2 \cos^2 \theta)^{1/2}
  \ ,
\end{equation}
and the free-fall time $t_\ff$ and
free-fall azimuthal angle $\phi_\ff$
are related to the usual Boyer-Lindquist time $t$ and azimuthal angle $\phi$ by
\begin{equation}
  t_\ff = t - \int_r^\infty {\vff \, \dd r \over 1 - \vff^2}
\end{equation}
\begin{equation}
  \phi_\ff = \phi - a \int_r^\infty {\vff \, \dd r \over R^2 (1 - \vff^2)}
  \ .
\end{equation}
As before,
we adopt the convention that the river velocity $\vff$
is positive for a black hole
(infalling),
negative for a white hole
(outfalling).
Horizons occur (see \S\ref{horizons})
where the river velocity $\vff$ equals the speed of light
\begin{equation}
\label{vhor}
  \vff = \pm 1
\end{equation}
with $\vff = 1$ for black hole horizons,
and $\vff = -1$ for white hole horizons.
Ergospheres occur where
$\dd s^2 = 0$ at $\dd r = \dd\theta = \dd\phi_\ff = 0$,
which happens at
\begin{equation}
  \vff = \pm {\rho \over R}
\end{equation}
again with $\vff = \rho / R$ for black hole ergospheres,
and $\vff = - \rho / R$ for white hole ergospheres.
For a Kerr-Newman black hole with mass $M$ and charge $Q$,
the river velocity $\vff$ is
\begin{equation}
\label{vffkerrnewman}
  \vff(r) = {( 2 M r - Q^2 )^{1/2} \over R}
\end{equation}
but for the present purpose
the river velocity can be considered to be a more general function of
the radial coordinate $r$.
Note that the river velocity $\vff$ defined here differs from
Doran's \cite{Doran} velocity by a factor of $\rho/R$.
Doran defines the velocity to equal the magnitude
$( \vff_\mu \vff^\mu )^{1/2} = \vff R / \rho$ of the
velocity vector $\vff^\mu$ given by equation~(\ref{vmu}) below,
a seemingly natural choice.
The point of the convention adopted here is that $\vff(r)$
is any and only a function of $r$,
rather than depending also on $\theta$
through $\rho \equiv (r^2 + a^2 \cos^2 \theta)^{1/2}$.
Moreover, with the convention here,
the river velocity is plus or minus one at horizons,
equation~(\ref{vhor}),
as will be demonstrated below, \S\ref{horizons}.

If the river velocity $\vff$ is zero,
then the metric~(\ref{doranmetric}) reduces to the flat space metric
in oblate spheroidal coordinates.
However, unlike the spherical case,
the metric is not flat along hypersurfaces of constant free-fall time,
$\dd t_\ff = 0$.

\subsection{Doran-Cartesian metric}
\label{dorancartesian}

The Doran coordinate system turns out, \S\S\ref{flatbkgd} and \ref{flatbkgdre},
to provide the coordinates of the flat background
through which the river of space
flows into the black hole.
We therefore express the Doran metric in Cartesian coordinates
$x^\mu = (x^0,x^1,x^2,x^3)$
=
$(t_\ff,x,y,z)$
=
$(t_\ff, R \sin\theta\cos\phi_\ff, R \sin\theta\sin\phi_\ff, r \cos\theta)$
with the rotation axis taken along the $z$-direction:
\begin{equation}
\label{dorancartesianmetric}
  \dd s^2
  =
  \eta_{\mu\nu}
  ( \dd x^\mu - \vff^\mu \alpha_\kappa \dd x^\kappa )
  ( \dd x^\nu - \vff^\nu \alpha_\lambda \dd x^\lambda )
  \ .
\end{equation}
Here the components of the river velocity $\vff^\mu$ are
\begin{equation}
\label{vmu}
  \vff^\mu
  = {\vff R \over \rho} \left(
    0 \, , \, - {x r \over R \rho} \, , \, - {y r \over R \rho} \, ,
    \, - {z R \over r \rho}
  \right)
\end{equation}
and
$\alpha_\mu \dd x^\mu = \dd t_\ff - a \sin^2\!\theta \, \dd\phi_\ff$
has components
\begin{equation}
\label{amu}
  \alpha_\mu
  =
  \left( 1 \, , \, {a y \over R^2} \, , \, - {a x \over R^2} \, , \, 0 \right)
  \ .
\end{equation}
The vector $\alpha_\mu$ is related to the 4-velocity of the horizon,
equation~(\ref{uhor}),
and we refer to it as
the azimuthal vector,
since its spatial components point in the (negative) azimuthal direction,
in the direction opposite to the rotation of the black hole.
The spheroidal radial coordinate $r$ is given implicitly
in terms of $x,y,z$ by
\begin{equation}
  r^4 - r^2 (x^2 + y^2 + z^2 - a^2) - a^2 z^2 = 0
  \ .
\end{equation}

\subsection{River tetrad}

In modelling black holes as an inflowing river of space,
it is natural to work in the orthonormal tetrad formalism.
Let $\gvec_\mu$ denote the basis of tangent vectors
in the Doran-Cartesian coordinate system $x^\mu$,
and let $\gammavec_m$,
and associated local coordinates $\xi^m = (\xi^0,\xi^1,\xi^2,\xi^3)$,
denote a system of locally inertial frames, tetrads,
attached to observers who free-fall from zero velocity
(with zero angular momentum) at infinity.
Such freely-falling observers
are comoving with the infalling river of space.
They fall along trajectories of constant $\theta$ and $\phi_\ff$,
and have 4-velocities
$\ucoord_\ff^\mu = (1, \vff^1, \vff^2, \vff^3)$
in the Doran-Cartesian coordinate system.
The scalar products of the tangent vectors $\gvec_\mu$ at each point
constitute the metric $g_{\mu\nu}$,
equation~(\ref{gg}),
while the scalar products of the tetrad vectors $\gammavec_m$ at each point
form the Minkowski metric,
equation~(\ref{gammaeta}).
If the tetrad frames $\gammavec_m$ are assumed, without loss of generality,
to be aligned with the tangent vectors $\gvec_\mu$ at infinity,
then the relation between $\gammavec_m$ and $\gvec_\mu$ is,
as previously noted by Doran (2000) \cite{Doran},
\begin{eqnarray}
\label{gammarot}
  \gammavec_{t_\ff}
  &=&
  \gvec_{t_\ff}
  + \vff^i \gvec_i
\nonumber
\\
  \gammavec_x
  &=&
  \gvec_x
  +
  \alpha_x \vff^i \gvec_i
\nonumber
\\
  \gammavec_y
  &=&
  \gvec_y
  +
  \alpha_y \vff^i \gvec_i
\\
  \gammavec_z
  &=&
  \gvec_z
\nonumber
\end{eqnarray}
which may be confirmed by checking that the scalar products of the $\gammavec_m$
so constructed form the Minkowski metric,
and that their derivatives vanish along the worldlines of observers
who free-fall from zero velocity at infinity,
$\ucoord_\ff^\mu \, \partial \gammavec_m / \partial x^\mu = 0$.
If horizontal radial and azimuthal axes are defined by
$(\gammavec_\rightarrow, \gammavec_\uparrow) \equiv
(\cos \phi_\ff \, \gammavec_x + \sin \phi_\ff \, \gammavec_y ,
\discretionary{}{}{}
- \sin \phi_\ff \, \gammavec_x + \cos \phi_\ff \, \gammavec_y)$
and likewise
$(\gvec_\rightarrow, \gvec_\uparrow) \equiv
(\cos \phi_\ff \, \gvec_x + \sin \phi_\ff \, \gvec_y ,
\discretionary{}{}{}
- \sin \phi_\ff \, \gvec_x + \cos \phi_\ff \, \gvec_y)$,
then
\begin{eqnarray}
\label{gammarperp}
  \gammavec_\rightarrow
  &=&
  \gvec_\rightarrow
\nonumber
\\
  \gammavec_\uparrow
  &=&
  \gvec_\uparrow
  -
  {a \sin\theta \over R}
  \vff^i \gvec_i
  \ .
\end{eqnarray}
Equations~(\ref{gammarot}) and (\ref{gammarperp})
show that the time axis
$\gammavec_{t_\ff}$
is shifted by velocity $\vff^i \gvec_i$,
similar to the spherical case,
equation~(\ref{gammaspher}),
but in addition the azimuthal axis $\gammavec_\uparrow$
is shifted by $- ( a \sin\theta / R ) \vff^i \gvec_i$.
Figure~\ref{tetrad} illustrates the
horizontal radial and azimuthal axes
$\gammavec_\rightarrow$ and $\gammavec_\uparrow$
at several points
in the equatorial plane of a Kerr black hole.
The azimuthal axes $\gammavec_\uparrow$ are tilted radially,
in accordance with equation~(\ref{gammarperp}),
reflecting the fact that the spatial metric is sheared.

\tetradfig

Equations~(\ref{gammarot})
may be abbreviated
$\gammavec_m = {e_m}^\mu \gvec_\mu$
where ${e_m}^\mu$ is the vierbein
\begin{equation}
\label{vierbein}
  {e_m}^\mu = \delta_m^\mu + \alpha_m \vff^\mu
\end{equation}
with $\delta_m^\mu$ a Kronecker delta.
The inverse vierbein
${e^m}_\mu$
is
\begin{equation}
\label{vierbeininverse}
  {e^m}_\mu = \delta_\mu^m - \alpha_\mu \vff^m
  \ .
\end{equation}
That the product of the vierbein and its inverse
given by equations~(\ref{vierbein}) and (\ref{vierbeininverse})
is indeed the unit matrix,
${e_m}^\mu {e^n}_\mu = \delta_m^n$ and ${e^m}_\mu {e_m}^\nu = \delta_\mu^\nu$,
follows from
the orthogonality of the azimuthal and velocity vectors
$\alpha_\mu$ and $\vff^\mu$,
namely $\alpha_\mu \vff^\mu = 0$.
The vectors
$\alpha_m$ with a latin index in the vierbein~(\ref{vierbein})
and $\vff^m$ with a latin index in the inverse vierbein~(\ref{vierbeininverse})
are defined by
\begin{equation}
\label{av}
  \alpha_m \equiv \delta_m^\mu \alpha_\mu
  \ ,
  \qquad
  \vff^m \equiv \delta^m_\mu \vff^\mu
\end{equation}
and transform with the tetrad frame $\gammavec_m$
rather than the coordinate frame $\gvec_\mu$.
The coordinates of $\alpha_m$ and $\vff^m$ are the same as those of
$\alpha_\mu$ and $\vff^\mu$
in the particular tetrad frame and coordinate system we are using,
but would be different in a different tetrad frame
or a different coordinate system.

In general,
the vierbein ${e_m}^\mu$ and its inverse ${e^m}_\mu$
provide the means of transforming the components
$p_\mu$ or $p^\mu$
of any arbitrary 4-vector
between the coordinate frame and the tretrad frame
\begin{equation}
\label{pm}
  p_m = {e_m}^\mu p_\mu
  \ ,
  \qquad
  p^m = {e^m}_\mu p^\mu
  \ .
\end{equation}
Indices on vectors $p_m$ and $p^m$
in the tetrad frame
are raised and lowered with the Minkowski metric $\eta_{mn}$,
whereas
indices on vectors $p_\mu$ and $p^\mu$
in the coordinate frame
are raised and lowered with the coordinate metric $g_{\mu\nu}$.

As a particular case of equations~(\ref{pm}),
it is true that
\begin{equation}
\label{ave}
  \alpha_m = {e_m}^\mu \alpha_\mu
  \ ,
  \qquad
  \vff^m = {e^m}_\mu \vff^\mu
\end{equation}
which reduces to the asserted definitions~(\ref{av})
thanks to the orthogonality of $\alpha_\mu$ and $\vff^\mu$.
If the coordinate system or tetrad frame is changed,
then the vierbein change accordingly,
and $\alpha_m$ and $\vff^m$
change in accordance with equations~(\ref{ave}).


The components $\utet^m$ of the 4-velocity of a particle
relative to the the tetrad frame
are related to the components $\ucoord^\mu$ in the coordinate frame by
$\utet^m = {e^m}_\mu \ucoord^\mu$,
or explicitly
\begin{eqnarray}
\label{gammau}
  \utet^0
  &=&
  \ucoord^0
\nonumber
\\
  \utet^i
  &=&
  \ucoord^i - \vff^i \alpha_\mu \ucoord^\mu
  \qquad
  (i=1,2,3)
  \ .
\end{eqnarray}
Equations~(\ref{gammau}) say that
if in an interval of proper time $\delta \tau$ the particle moves
a coordinate distance $\delta x^\mu = \ucoord^\mu \delta \tau$,
then relative to the tetrad frame,
that is,
relative to the locally inertial frame of an observer who is comoving with the
infalling river,
the particle moves
a proper distance
\begin{equation}
\label{dxi}
  \delta \xi^m
  = {e^m}_\mu \delta x^\mu
  = \delta x^m - \vff^m \alpha_\mu \delta x^\mu
  \ .
\end{equation}
One recognizes the right hand side of equation~(\ref{dxi})
as having the same form as a factor of the Doran-Cartesian
metric~(\ref{dorancartesianmetric}).
The temporal displacement $\delta \xi^0$ of the particle in the tetrad frame
is the Galilean time change $\delta t_\ff$,
as in the spherical case.
However,
the proper spatial displacement $\delta \xi^i$
of the particle in the tetrad frame
differs from the displacement $\delta x^i$ in the coordinate frame
not by the Galilean distance $\vff^i \delta t_\ff$
that the river moves in time $\delta t_\ff$,
as in the spherical case,
but rather by
$\vff^i \alpha_\mu \delta x^\mu
= \vff^i (\delta t_\ff - a \sin^2\!\theta \, \delta \phi_\ff)$.
The extra part $- \vff^i a \sin^2\!\theta \, \delta \phi_\ff$
arises from the spatial shear in the metric,
illustrated in Figure~\ref{tetrad}.

\subsection{Horizons}
\label{horizons}

It is now possible to see how the position of horizons
is set by $\vff = \pm 1$, as earlier asserted,
equation~(\ref{vhor}).
It follows from the previous paragraph
that the effective velocity of the river,
from the point of view of an object in the river,
depends on the state of motion of the object.
The effective river velocity is
$\vff^i \alpha_\mu \dd x^\mu / \dd t_\ff$,
which differs from $\vff^i$ by the factor
$\alpha_\mu \ucoord^\mu / \ucoord^0
= \alpha_\mu \dd x^\mu / \dd t_\ff
= 1 - a \sin^2\!\theta \, \dd\phi_\ff / \dd t_\ff$.
Irrespective of this factor,
the effective river velocity is always pointed radially
(along lines of constant $\theta$ and $\phi_\ff$)
inward along the direction of $\vff^i$.
If we restrict temporarily to considering
only objects with a given fixed value of
$\alpha_\mu \ucoord^\mu / \ucoord^0$,
then such objects can escape outward only if their radial velocity
\begin{equation}
  \ucoord^r = {\partial r \over \partial x^\mu} \ucoord^\mu
\end{equation}
exceeds zero.
To determine the position of the horizon,
we may thus first solve the slightly more general problem
of maximizing the radial velocity $\ucoord^r$
subject to constraints on $\ucoord^0$
(which can be set to $1$ without loss of generality),
$\alpha_\mu \ucoord^\mu$,
and $\ucoord_\mu \ucoord^\mu$,
the last constraint coming from the fact
that the 4-velocity must be time-like or light-like,
requiring $\ucoord_\mu \ucoord^\mu \leq 0$.
Equivalently,
we can minimize
$\ucoord_\mu \ucoord^\mu$
subject to constraints on $\ucoord^0$, $\alpha_\mu \ucoord^\mu$,
and $\ucoord^r$.
This implies that the 4-velocity must satisfy
\begin{equation}
  \ucoord_\mu
  =
  \lambda \delta^0_\mu + \mu \alpha_\mu + \nu {\partial r \over \partial x^\mu}
\end{equation}
where $\lambda,\mu,\nu$ are Lagrange multipliers,
whose values are determined by fixing any three of the four quantities
$\ucoord^0$, $\ucoord^r$, $\alpha_\mu \ucoord^\mu$,
and $\ucoord_\mu \ucoord^\mu$.
Not surprisingly,
the largest value of $\ucoord^r$ at fixed $\ucoord^0$
and $\alpha_\mu \ucoord^\mu$
occurs when the 4-velocity is light-like, $\ucoord_\mu \ucoord^\mu = 0$.
Eliminating the Lagrange multipliers $\lambda,\mu,\nu$
in favor of
$\ucoord^0 = 1$, $\ucoord^r = 0$, and $\ucoord_\mu \ucoord^\mu = 0$,
yields
\begin{equation}
  {\alpha_\mu \ucoord^\mu \over \ucoord^0}
  =
  {\rho^2 \left[ R \pm a \sin\theta (1 - \vff^2)^{1/2} \right]
  \over R ( \rho^2 + \vff^2 a^2 \sin^2\!\theta )}
\end{equation}
which has a real solution provided that $\vff^2 \leq 1$,
with $\alpha_\mu \ucoord^\mu / \ucoord^0 = \rho^2 / R^2$ at $\vff^2 = 1$.
The position of the horizon is thus set by $\vff^2 = 1$, as claimed:
if $\vff^2 < 1$, then there are geodesics on which a particle can escape,
$\ucoord^r > 0$;
if on the other hand $\vff^2 > 1$, then all geodesics are trapped,
and an object is compelled to fall inward
(or outward, in the case of a white hole).

The 4-velocity of a photon that just holds steady on the horizon,
a member of the outgoing principal null congruence,
satisfies
$\ucoord_\mu = (\rho^2 / R^2) \, \partial r / \partial x^\mu$,
and is
\begin{equation}
\label{uhor}
  \ucoord^\mu =
  \left( 1 \, , \, - {a y \over R^2} \, , \, {a x \over R^2} \, , \, 0 \right)
  \ .
\end{equation}
Interestingly,
the contravariant components $\ucoord^\mu$ of this 4-velocity
coincide, modulo a minus sign,
with the covariant components $\alpha_\mu$ of the azimuthal vector,
equation~(\ref{amu}).
Relative to the river frame,
the horizon rotates right-handedly with angular velocity
\begin{equation}
  {\dd \phi_\ff \over \dd t_\ff} = {a \over R^2}
\end{equation}
which is also the angular velocity of the horizon
perceived by an observer at rest at infinity.

\subsection{Equations of motion in the tetrad formalism}
\label{eqmottetrad}

Our aim in this subsection
is to derive equations of motion for objects
relative to the inflowing river of space.
For clarity and pedagogy,
we start from basic principles
to derive the equations of motion~(\ref{dpdtau})
of 4-vectors in the tetrad frame.
Having derived the equations of motion~(\ref{dpdtau}),
we will describe what these equations mean physically.
In the next subsection, \S\ref{flatbkgd},
we will go on to apply these equations to the particular case of black holes,
where the vierbein are given by equation~(\ref{vierbein}).

Let
$\pvec$
be an arbitrary 4-vector.
The 4-vector
$\pvec = p^m \gammavec_m = p^\mu \gvec_\mu$
is an invariant object,
independent of the choice of tetrad or coordinate system.
According to the Principle of Equivalence,
an unaccelerated 4-vector $\pvec$ remains at rest in its own free-fall frame,
meaning that its derivative with respect to its own proper time $\tau$ is zero
in its own frame
\begin{equation}
\label{dpvecdtau}
  {\dd \pvec \over \dd  \tau} = 0
  \ .
\end{equation}
If the 4-vector $\pvec$ is experiencing an acceleration
in its own frame
(perhaps because of an electromagnetic field,
or perhaps because of rockets being fired),
then the zero on the right hand side of equation~(\ref{dpvecdtau})
should be replaced by an appropriate invariant acceleration 4-vector.
Here we set any such acceleration to zero,
recognizing that an acceleration could be reinstated if desired
at the end of the calculation.
Since $\pvec$ is invariant,
equation~(\ref{dpvecdtau}) must be true in all frames.
In the tetrad frame, this implies
\begin{equation}
\label{gdpdtau}
  \gammavec_m {\dd p^m \over \dd \tau} + {\dd  \gammavec_m \over \dd  \tau} p^m
  =
  0
  \ .
\end{equation}
The proper time derivative $\dd / \dd  \tau$ can be written
\begin{equation}
\label{ddtau}
  {\dd \over \dd  \tau}
  = \ucoord^\nu {\partial \over \partial x^\nu}
  = \utet^n {e_n}^\nu {\partial \over \partial x^\nu}
  = \utet^n \partial_n
\end{equation}
where the directed derivative $\partial_n$ is defined by
\begin{equation}
\label{gammaderiv}
  \partial_n \equiv {e_n}^\nu {\partial \over \partial x^\nu}
  \ .
\end{equation}
The derivative $\partial_n$ defined by equation~(\ref{gammaderiv})
is independent of the choice of coordinates $x^\nu$,
as suggested by the absence of any greek index.
The derivative may be written
$\partial_n = \gammavec_n \cdot \partialvec$
where
$\partialvec
\equiv \gvec^\nu \, \partial / \partial x^\nu$
and
$\gvec^\nu \equiv g^{\nu\mu} \gvec_\mu$,
which shows that $\partial_n$ is a directed derivative along $\gammavec_n$,
the dot product of
the vector $\gammavec_n$
with
the vector derivative
$\partialvec \equiv
\gvec^\nu \, \partial / \partial x^\nu$,
a coordinate-independent object.
In other words,
$\partial_n$
constitute the tetrad frame components of the invariant 4-vector derivative
$\partialvec
= \gammavec^n \, \partial_n
= \gvec^\nu \, \partial / \partial x^\nu$.
Unlike the partial derivatives $\partial / \partial x^\nu$,
the directed derivatives $\partial_n$ do not commute.
In terms of the vierbein derivatives
$\de_{kmn}$
defined by
\begin{equation}
\label{gammad}
  \de_{kmn}
  \equiv
  \eta_{kl} \, {e^l}_\lambda \, {e_n}^\nu \,
  {\partial {e_m}^\lambda \over \partial x^\nu}
\end{equation}
the commutator
$\left[ \partial_k, \partial_m \right]$
of two directed derivatives is
\begin{equation}
\label{gammadcommutator}
  \left[ \partial_k, \partial_m \right]
  =
  {f_{km}}^n \partial_n
  \ ,
  \qquad
  f_{kmn} \equiv
  \de_{nmk} - \de_{nkm}
  \ .
\end{equation}
The $f_{kmn}$ are the structure coefficients
of the commutators of directed derivatives.

Introduce the tetrad frame connection coefficients $\Gamma^k_{mn}$,
also known as the Ricci rotation coefficients,
defined by
\begin{equation}
\label{gammaGammadef}
  \partial_n \gammavec_m \equiv \Gamma^k_{mn} \gammavec_k
  \ .
\end{equation}
In terms of the vierbein ${e_m}^\mu$ and basis vectors $\gvec_\mu$,
the tetrad frame connection coefficients
with all indices lowered,
$\Gamma_{kmn} \equiv \eta_{kl} \Gamma^l_{mn}$,
are, from equation~(\ref{gammaGammadef}),
\begin{equation}
\label{gammaGammag}
  \Gamma_{kmn}
  = \gammavec_k \cdot \partial_n \gammavec_m
  = {e_k}^\kappa \gvec_\kappa \cdot {e_n}^\nu
  {\partial ({e_m}^\mu \gvec_\mu) \over \partial x^\nu}
  \ .
\end{equation}
The usual coordinate frame connection coefficients,
the Christoffel symbols
$\Gamma_{\kappa\mu\nu} \equiv g_{\kappa\lambda} \Gamma^\lambda_{\mu\nu}$,
are defined by
\begin{equation}
\label{Gammadef}
  {\partial \gvec_\mu \over \partial x^\nu}
  \equiv
  \Gamma^\kappa_{\mu\nu} \gvec_\kappa
  \ .
\end{equation}
Equations~(\ref{gammaGammag}) and (\ref{Gammadef})
imply that the tetrad frame connection coefficients
$\Gamma_{kmn}$
are related to the Christoffel symbols
$\Gamma_{\kappa\mu\nu}$
by
\begin{equation}
\label{GammaGamma}
  \Gamma_{kmn}
  =
  \de_{kmn}
  +
  {e_k}^\kappa {e_m}^\mu {e_n}^\nu
  \Gamma_{\kappa\mu\nu}
  \ .
\end{equation}
The definition~(\ref{gammaGammadef})
and the fact that
$\partial_n ( \gammavec_k \cdot \gammavec_m ) = \partial_n \eta_{km} = 0$
implies that the tetrad frame connection coefficients
$\Gamma_{kmn}$
are antisymmetric in their first two indices,
\begin{equation}
\label{Gammaantisym}
  \Gamma_{kmn} = - \Gamma_{mkn}
  \ .
\end{equation}
The tangent vectors
$\gvec_\mu$
can be regarded as coordinate derivatives of the invariant 4-vector interval
$d \xvec \equiv \gvec_\mu d x^\mu$,
that is,
$\gvec_\mu = \partial \xvec / \partial x^\mu$,
and the commutativity of partial derivatives,
$\partial \gvec_\mu / \partial x^\nu
=
\partial^2 \xvec / \partial x^\nu \partial x^\mu
=
\partial^2 \xvec / \partial x^\mu \partial x^\nu
=
\partial \gvec_\nu / \partial x^\mu$,
implies that the Christoffel symbols
$\Gamma^\kappa_{\mu\nu}$
are symmetric in their last two indices,
\begin{equation}
\label{notorsion}
  \Gamma^\kappa_{\mu\nu} =
  \Gamma^\kappa_{\nu\mu}
\end{equation}
which is the usual no-torsion condition of general relativity.
Combining
equation~(\ref{GammaGamma})
with the antisymmetry relation~(\ref{Gammaantisym})
and the no-torsion condition~(\ref{notorsion})
yields an expression for the tetrad frame connection coefficients
entirely in terms of the vierbein derivatives
$\de_{kmn}$
\begin{equation}
\label{gammaGamma}
  \Gamma_{kmn}
  =
  \frac{1}{2} \,
  \left(
  \de_{kmn} - \de_{mkn} + \de_{nmk} - \de_{nkm} + \de_{mnk} - \de_{knm}
  \right)
  \ .
\end{equation}

From equations~(\ref{gdpdtau}), (\ref{ddtau}) and (\ref{gammaGammadef})
it follows that the equations of motion for the tetrad components $p^k$
of an unaccelerated 4-vector $\pvec = p^k \gammavec_k$ are
\begin{equation}
\label{dpdtau}
  {\dd p^k \over \dd \tau} + \Gamma^k_{mn} \utet^n p^m = 0
  \ .
\end{equation}
The physical significance of the equations of motion~(\ref{dpdtau})
is as follows.
The tetrad $\gammavec_m$ defines a set of locally inertial
frames throughout spacetime.
In the present case,
these locally inertial frames have been constructed so that
an observer who free-falls from zero velocity at infinity
finds their own frame aligned at all times with the tetrad frame.
But in general another observer who free-falls along a different
geodesic will find their own locally inertial frame becoming misaligned
with the tetrad frame.
Equation~(\ref{dpdtau})
expresses this misalignment of locally inertial frames.
Because the misalignment is between locally inertial frames,
it is a Lorentz transformation.
This Lorentz transformation is encoded in the connections
$\Gamma^k_{mn}$.
Specifically,
if a 4-vector $p^k$ is transported in free-fall
by an infinitesimal distance $\delta \xi^n = \utet^n \delta \tau$
relative to the tetrad frame $\gammavec_n$,
then the 4-vector experiences an infinitesimal Lorentz transformation
$p^k \rightarrow p^k - \delta \xi^n \Gamma^k_{mn} p^m$.
In other words,
the connection coefficients $\Gamma^k_{mn}$
for each final index $n$
is the generator of a Lorentz transformation.

The antisymmetry of the tetrad frame connection coefficient
with respect to its first two indices,
equation~(\ref{Gammaantisym}),
expresses mathematically
the fact that $\Gamma^k_{mn}$ for each given $n$
is the generator of a Lorentz transformation.
Components of $\Gamma^k_{mn}$
in which one of the first two indices $k$ or $m$ is 0 (time)
generate Lorentz boosts.
Components of $\Gamma^k_{mn}$
in which both of the first two indices $k$ and $m$ are 1, 2, or 3 (space)
generate spatial rotations.

\subsection{The flat background}
\label{flatbkgd}

The previous subsection, \S\ref{eqmottetrad},
considered the equations of motion in the tetrad formalism in the general case.
We now particularize to the case at hand,
that of rotating black holes,
where the vierbein are given by equation~(\ref{vierbein}).
In this subsection
we see how the Doran-Cartesian coordinate system
emerges as the coordinate system of a flat background.
In the next subsection, \S\ref{riverfield},
we will see how the connection coefficients
are expressible as the flat space gradient of a river field.
In \S\ref{flatbkgdre},
we will revisit the notion of the flat background and what it means.

Explicit computation of the connection coefficients,
equation~(\ref{gammaGamma}),
from the vierbein of equation~(\ref{vierbein})
reveals that
the sea of terms nonlinear in the vierbeins
undergo a remarkable cancellation
(this is not just the Jacobi identity at work)
leaving only terms linear in the vierbeins ${e_m}^\lambda$.
In other words,
the connection coefficients reduce to
the same expression as~(\ref{gammaGamma}), but with the
${e^l}_\lambda \, {e_n}^\nu$
factors
in equation~(\ref{gammad}) for $\de_{kmn}$
replaced by Kronecker deltas
$\delta^l_\lambda \, \delta_n^\nu$
\begin{equation}
\label{gammadd}
  \de_{kmn}
  \rightarrow
  \eta_{kl} \, \delta^l_\lambda \, \delta_n^\nu \,
  {\partial {e_m}^\lambda \over \partial x^\nu}
  =
  \delta_n^\nu \, {\partial \alpha_m \vff_k \over \partial x^\nu}
  \ .
\end{equation}
The fact that the derivative
${e_n}^\nu \, \partial / \partial x^\nu$ in equation~(\ref{gammad})
gets replaced by $\delta_n^\nu \, \partial / \partial x^\nu$
in equation~(\ref{gammadd})
motivates introducing a new set of flat space coordinates $x^n$,
with latin indices,
with the defining property
that in the particular coordinate and tetrad frame that we are using
\begin{equation}
\label{partialxn}
  {\partial \over \partial x^n} \equiv
  \delta_n^\nu {\partial \over \partial x^\nu}
  \ .
\end{equation}
The invariant relation
$\dd x^n \, \partial / \partial x^n = \dd x^\nu \, \partial / \partial x^\nu$
then implies that the flat space differentials $\dd x^n$
are related to the coordinate differentials $\dd x^\nu$ by
\begin{equation}
\label{dxn}
  \dd x^n = \delta_\nu^n \, \dd x^\nu
  \ .
\end{equation}
It should be emphasized that the relations~(\ref{partialxn}) and (\ref{dxn})
should be interpreted as being true only in the particular tetrad
and coordinate frame that we are using.
If the tetrad frame is subjected to a local gauge transformation
(i.e.\ a Lorentz transformation that varies from place to place)
that rotates the locally inertial coordinates at each point by
$\xi^n \rightarrow \xi^{\prime \, n}$,
and if the coordinate system is subjected to a general coordinate transformation
$x^\nu \rightarrow x^{\prime \, \nu}$,
then the Kronecker deltas in equations~(\ref{partialxn}) and (\ref{dxn})
should be replaced by
\begin{equation}
\label{deltatransform}
  \delta_n^\nu \rightarrow
  \delta_m^\mu
  \,
  {\partial \xi^m \over \partial \xi^{\prime \, n}}
  {\partial x^{\prime \, \nu} \over \partial x^\mu}
  \ ,
  \qquad
  \delta_\nu^n \rightarrow
  \delta_\mu^m
  \,
  {\partial \xi^{\prime \, n} \over \partial \xi^m}
  {\partial x^\mu \over \partial x^{\prime \, \nu}}
  \ .
\end{equation}
In the particular tetrad and coordinate frame that we are using,
integrating the relation~(\ref{dxn}) arbitrarily through space yields
(the constant of integration being set to zero without loss of generality)
\begin{equation}
\label{xn}
  x^n = \delta_\nu^n \, x^\nu
  \ .
\end{equation}
Notwithstanding the index notation,
neither $x^n$ nor $x^\nu$ is a 4-vector
either under
local gauge transformations of the tetrad
or under
general transformations of the coordinates
(only the differentials $\dd x^n$ and $\dd x^\nu$ are 4-vectors),
so equation~(\ref{xn})
cannot be interpreted as a covariant equation relating
the coordinates $x^n$ and $x^\nu$,
even if the Kronecker delta is replaced according to
equations~(\ref{deltatransform}).
Rather, equation~(\ref{xn}) should be interpreted as true
in the particular coordinate and tetrad frame that we are using.
Equation~(\ref{xn})
can be regarded as defining the flat space coordinates $x^n$:
they are numerically the same as the curved space coordinates $x^\nu$
of the Doran-Cartesian metric~(\ref{dorancartesianmetric}),
but reincarnated as coordinates $x^n$
of a flat space with a Minkowski metric.
The Doran \cite{Doran} coordinate system
thus emerges as a rather special one,
providing the coordinates of the flat background
through which the river of space flows in rotating black holes.

The flat spacetime coordinates $x^n$
are not the same as the locally inertial coordinates $\xi^n$
attached to the tetrad $\gammavec_n$ at each point of spacetime.
The locally inertial differentials $\dd\xi^n$
are related to the coordinate differentials $\dd x^\nu$ by
\begin{equation}
  \dd\xi^n = {e^n}_\nu \, \dd x^\nu
\end{equation}
which differs from corresponding relation~(\ref{dxn}) between
$\dd x^n$ and
$\dd x^\nu$.

\subsection{The river field}
\label{riverfield}

The vectors $\alpha_m$ and $\vff^m$ can be regarded as functions
of the flat space coordinates $x^n$,
and the replacement of the vierbein derivatives $\de_{kmn}$,
equation~(\ref{gammadd}),
in the connection coefficients
can be written
\begin{equation}
  \de_{kmn} \rightarrow {\partial \alpha_m \vff_k \over \partial x^n}
  \ .
\end{equation}
The connection coefficients,
equation~(\ref{gammaGamma}),
are then given by flat space derivatives of $\alpha_m$ and $\vff_m$
\begin{eqnarray}
\label{gammaGammaav}
  \Gamma_{kmn}
  &=&
  \frac{1}{2}
  \left(
  {\partial \alpha_m \vff_k \over \partial x^n}
  -
  {\partial \alpha_k \vff_m \over \partial x^n}
  +
  {\partial \alpha_m \vff_n \over \partial x^k}
  -
  {\partial \alpha_n \vff_k \over \partial x^m}
  \right.
\nonumber
\\
  &&
  \left.
  + \,
  {\partial \alpha_n \vff_m \over \partial x^k}
  -
  {\partial \alpha_k \vff_n \over \partial x^m}
  \right)
  \ .
\end{eqnarray}

The connection coefficients with zero final index $n = 0$ are all
identically zero, $\Gamma_{km0} = 0$,
and taking the spatial curl of $\Gamma_{kmn}$ on the $n$ index
yields another sea of terms which again
undergo a remarkable cancellation
to nothing
\begin{equation}
\label{curlGamma}
  \varepsilon^{ijn} {\partial \Gamma_{kmn} \over \partial x^j}
  = 0
\end{equation}
for all $i,k,m$.
This demonstrates that the connection coefficients $\Gamma_{kmn}$
must be expressible as (minus) the flat space gradient $\partial / \partial x^n$
of an object $\omega_{km}$,
which we call the river field
since it encapsulates all the properties of the river
in the river model:
\begin{equation}
\label{gammaGammaomega}
  \Gamma_{kmn} = - {\partial \omega_{km} \over \partial x^n}
  \ .
\end{equation}
The river field $\omega_{km}$ is a bivector \cite{MTW,GLD},
inheriting from $\Gamma_{kmn}$ the property of being antisymmetric in $km$.
That the connection coefficient $\Gamma_{kmn}$ is the flat space gradient
of the river field lies at the heart of the river model
as a description of black holes.
After some manipulation we find the desired bivector river field to be
\begin{equation}
\label{omega}
  \omega_{km}
  =
  \alpha_k \vff_m - \alpha_m \vff_k + \varepsilon_{0kmi} \, \zeta^i
\end{equation}
where the vector $\zeta^i$ is
\begin{equation}
\label{zeta}
  \zeta^i = \left( 0 , 0 , 0 , \zeta \right)
  \ ,
  \quad
  \zeta = a \int_r^\infty {\vff \, \dd r \over R^2}
\end{equation}
which points vertically upward along the rotation axis of the black hole.

\kerrfig

The river field $\omega_{km}$
given by equation~(\ref{omega})
inherits from the connection coefficient $\Gamma_{kmn}$
its Lorentz structure.
The river field defines a velocity and a rotation, or twist,
at each point of the black hole geometry.
Components of $\omega_{km}$ in which one of the indices $k$ or $m$ is 0 (time)
define a velocity,
while components in which both indices $k$ and $m$ are $1,2,3$ (space)
define a spatial rotation, or twist.
The velocity is just the river velocity $\vff_m$
\begin{equation}
\label{omegatime}
  \omega_{0m} = \vff_m
  \ ,
\end{equation}
while the angle and axis of the river twist are given by the rotation vector
\begin{equation}
\label{omegaspace}
  \mu^i =
  {1 \over 2} \, \varepsilon^{ikm} \, \omega_{km} =
  \varepsilon^{ikm} \, \alpha_k \vff_m + \zeta^i
  \qquad
  (i,k,m = 1,2,3)
  \ .
\end{equation}
Like the velocity vector $\vff_i$,
the twist vector $\mu^i$ at each point lies in the plane
of constant free-fall azimuthal angle $\phi_\ff$,
since it is a sum of two vectors
$\varepsilon^{ikm} \, \alpha_k \vff_m$ and $\zeta^i$
both of which are orthogonal to the azimuthal vector $\alpha_k$.


Figure~\ref{kerr}
illustrates the velocity and twist fields
$\vff_i$ and $\mu_i$
for an uncharged black hole with specific angular momentum $a = 0.96$.

Another familiar bivector is the electromagnetic field tensor $F_{km}$,
and it can be useful to think of the river field bivector $\omega_{km}$
in those terms.
The velocity vector $\vff_i$ is the analog of the electric field vector $E_i$,
while the twist vector $\mu_i$ is the analog of the magnetic field vector $B_i$.
The analogy extends to the fact that, like a static electric field,
the velocity vector $\vff_i$ is the gradient of a potential $\psi$,
\begin{equation}
  \vff_i = - {\partial \psi \over \partial x^i}
  \ ,
  \quad
  \psi \equiv - \int_r^\infty \vff \, \dd r
  \ .
\end{equation}
However, unlike a magnetic field,
the twist vector $\mu^i$ is not pure curl,
although curiously $\mu^i + \zeta^i$ is pure curl,
having zero divergence,
$\partial (\mu^i + \zeta^i ) / \partial x^i = 0$.

\subsection{Motion of objects in the river}
\label{motion}

We are now ready to demonstrate a fundamental feature of the river model
for stationary rotating black holes,
that as an object moves through the river of space,
it is Lorentz boosted and rotated
by the tidal gradients in the velocity and twist fields of the river.

It follows from inserting the connection coefficients $\Gamma_{kmn}$
from equation~(\ref{gammaGammaomega}) into the equation of motion~(\ref{dpdtau})
that the equation of motion of an unaccelerated 4-vector $p^k$
in the river frame is
\begin{equation}
\label{dprot}
  {\dd p^k \over \dd \tau}
  =
  {\partial {\omega^k}_m \over \partial x^n} \utet^n p^m
  \ .
\end{equation}
The equation of motion~(\ref{dprot})
can be interpreted as follows.
In an infinitesimal interval $\delta \tau$ of proper time,
a particle moves a distance
$\delta \xi^n = \utet^n \delta \tau$
relative to the infalling river of space.
As a result of its motion through the river,
the particle experiences a tidal change
\begin{equation}
\label{domega}
  \delta {\omega^k}_m = {\partial {\omega^k}_m \over \partial x^n} \delta \xi^n
\end{equation}
in the river field,
which generalizes equation~(\ref{dv}) for spherical black holes.
The tidal change $\delta {\omega^k}_m$ in the river field
is an infinitesimal Lorentz transformation,
and it induces a Lorentz boost and rotation in the 4-vector $p^k$
\begin{equation}
\label{dpomega}
  p^k \rightarrow p^k + \delta {\omega^k}_m \, p^m
  \ .
\end{equation}
Equations~(\ref{domega}) and (\ref{dpomega})
reproduce the equations of motion~(\ref{dprot}).

\geodesicsfig

Figure~\ref{geodesics}
shows two ensembles of geodesics computed
using the equation of motion~(\ref{dprot}).
Each ensemble consists of a central point
and associated tetrad axes
surrounded by a set of points that are initially uniformly spaced around,
and initially at rest relative to, the central point,
in the locally inertial frame of the central point.
In the upper ensemble,
the central point is comoving with the infalling river of space,
while in the lower ensemble
the central point is initially moving radially outward,
but soon turns around and falls inward.
The tetrad axes are skewed because the spatial metric is sheared
(compare Figure~\ref{tetrad}).
In the lower ensemble the tetrad axes are
Lorentz-contracted in the radial direction because of the initial outward motion
of the ensemble relative to the infalling river.
The ensembles of points become tidally distorted
as they fall into the black hole.
If the locally inertial coordinates of a tetrad axis
are denoted $\delta \xi^k$,
then the tetrad axis
evolves according to
the equation of motion~(\ref{dprot}) with $p^k \rightarrow \delta \xi^k$,
\begin{equation}
\label{dxidtau}
  {\dd \delta \xi^k \over \dd \tau}
  =
  {\partial {\omega^k}_m \over \partial x^n} \utet^n \delta \xi^m
  \ .
\end{equation}
Similarly,
the tetrad 4-velocity $\utet^k$ of each point in the ensemble
evolves according to
the equation of motion~(\ref{dprot}) with $p^k \rightarrow \utet^k$,
\begin{equation}
\label{dudtau}
  {\dd \utet^k \over \dd \tau}
  =
  {\partial {\omega^k}_m \over \partial x^n} \utet^n \utet^m
  \ .
\end{equation}
Each point surrounding the central point
is initially at rest relative to the central point,
in the latter's locally inertial frame.
This requires that the covariant difference in tetrad 4-velocities
between each point and the central point initially vanishes,
which requires that the difference $\delta \utet^k$ in the tetrad 4-velocity
of a point separated from the central point by locally inertial separation
$\delta \xi^k$
initially satisfies,
to linear order in the separation $\delta \xi^k$,
\begin{equation}
\label{duomega}
  \delta \utet^k
  =
  {\partial {\omega^k}_m \over \partial x^n} \delta \xi^n \utet^m
  \ .
\end{equation}
The difference $\delta \utet^k$ in equation~(\ref{duomega})
is to be understood as
the tetrad 4-velocity of a point evaluated in the tetrad frame at that point,
minus the tetrad 4-velocity of the central point
evaluated in the tetrad frame at the central point.
Notice that the indices on $\delta \xi^n$ and $\utet^m$
on the right hand side of equation~(\ref{duomega})
are swapped compared to those
on the right hand side of equation~(\ref{dxidtau}).
In equation~(\ref{dxidtau})
the axis $\delta \xi^k$ is transported along the 4-velocity $\utet^n$,
whereas in equation~(\ref{duomega})
the 4-velocity $\utet^k$ is transported along the axis $\delta \xi^n$.

The lower ensemble of points in Figure~\ref{geodesics}
illustrates the twist in the locally inertial frame
that develops as the ensemble moves through the river of space.
The twist acts so as to keep the locally inertial frame
comoving with the geodesic motion of points
in a small neighborhood of the frame.

Equation~(\ref{duomega}) is true initially,
when the ensemble of points are at rest relative to each other,
satisfying $D \delta \xi^k / D \tau = 0$.
The more general form of equation~(\ref{duomega}),
valid when the points are in relative motion is,
to linear order in the separation $\delta \xi^k$
(here we revert, eq.~(\ref{gammaGammaomega}),
to the more familiar notation for the connections
$\Gamma^k_{mn}$),
\begin{equation}
  \delta \utet^k + \Gamma^k_{mn} \delta \xi^n \utet^m
  =
  {D \delta \xi^k \over D \tau}
\end{equation}
or equivalently
\begin{equation}
\label{duomegax}
  \delta \utet^k =
  {\dd \delta \xi^k \over \dd \tau}
  + \left( \Gamma^k_{nm} - \Gamma^k_{mn} \right) \delta \xi^n \utet^m
  \ .
\end{equation}
Variation of the equation of motion~(\ref{dudtau}) for $\utet^k$ gives
\begin{equation}
\label{ddudtau}
  {\dd \delta \utet^k \over \dd \tau}
  + \partial_l \Gamma^k_{mn} \delta \xi^l \utet^n \utet^m
  + \Gamma^k_{mn} \left( \utet^n \delta \utet^m + \utet^m \delta \utet^n \right)
  = 0
  \ .
\end{equation}
Inserting the expression~(\ref{duomegax}) for $\delta \utet^k$
into this equation~(\ref{ddudtau}) yields the familiar equation
of geodesic deviation
\begin{equation}
  {D^2 \delta \xi^k \over D \tau^2}
  =
  {R_{lmn}}^k \utet^m \utet^n \delta \xi^l
\end{equation}
where
$R_{klmn}$
is the Riemann curvature tensor
\begin{eqnarray}
  R_{klmn}
  &=&
  \partial_l \Gamma_{nmk} - \partial_k \Gamma_{nml}
  + \Gamma^j_{ml} \Gamma_{jnk} - \Gamma^j_{mk} \Gamma_{jnl}
\nonumber
\\
  &&
  + \, ( \Gamma^j_{lk} - \Gamma^j_{kl} ) \Gamma_{nmj}
  \ .
\end{eqnarray}


%
%

\subsection{The flat background revisited}
\label{flatbkgdre}

Now that we have completed the formalism of the river model,
it is useful to revisit the question of the flat background, \S\ref{flatbkgd},
through which the river of space flows and twists into a rotating black hole.
What exactly does flatness mean in this context?

The crucial equation is equation~(\ref{gammaGammaomega}),
which states that the connection coefficient is given by the flat space
gradient of the river field.
The fact that the gradient is an ordinary partial derivative
with respect to Doran-Cartesian coordinates
is what makes the background flat,
and in a sense that is all there is to it.
Equation~(\ref{gammaGammaomega})
acquires physical significance because it
propagates through to the equation of motion~(\ref{dprot})
of objects swimming in the river.
The equation of motion paints the physical picture
of objects moving in the river being Lorentz boosted and rotated
by the flat space tidal gradients in the velocity and twist components
of the river field.

The statement that the background spacetime in the river model is flat
is not a statement about the metric $g_{\mu\nu}$ being flat.
Rulers and clocks swimming in the river of space measure
not distances and times in the background space,
but rather distances and times
relative to the tidally twisting and stretching river.
The presence of tides is the signature of curvature,
so it makes sense that the metric measured by rulers and clocks is not flat.

It is to be emphasized that
the flat background has no physically observable meaning.
It is simply a fictitious construct
that somehow emerges from the mathematics.

\section{Summary}
\label{summary}

In this paper we have presented a
way to conceptualize stationary black holes,
which we call the river model.
The river model offers a mental picture of black holes
which is intuitively appealing,
and whose basic elements are simple enough
that they can be grasped by non-experts.
In the river model, space itself flows like a river through a flat background,
while objects move through the river
according to the rules of special relativity.
For a Schwarzschild (non-rotating, uncharged) black hole,
the river falls radially inward at the Newtonian escape velocity,
hitting the speed of light at the horizon.
Inside horizons, the river of space moves faster than light,
carrying everything with it.

We have presented the details
that place the river model on a sound mathematical basis.
We have shown that the river model works for any stationary black hole,
rotating as well as non-rotating,
charged as well as uncharged.
The Doran \cite{Doran} coordinate system provides the coordinates of
the flat background through which the river of space flows into the black hole.

The extension of the river model to rotating black holes
proves to be both surprising and pretty.
Contrary to expectation, the river does not spiral into a rotating black hole:
the azimuthal component of the river velocity is zero.
Instead,
the river has at each point not only a velocity,
but also a rotation, or twist.
The river is thus a Lorentz river,
characterized by all six generators of the Lorentz group.
As an object moves through the river of space,
it is Lorentz boosted by changes in the velocity of the river along its path,
and rotated by changes in the twist of the river.
Equation~(\ref{omega}) gives an explicit expression for the river field,
a six-component bivector field that specifies the velocity and twist
of the river at each point of the black hole geometry.

The tidal boosts and twists experienced by an object in the river
induce a curvature in the spacetime measured by the object,
causing the metric to be non-flat.
Changes in the river velocity rotate between space and time axes,
while changes in the river twist rotate between two spatial axes.
For a spherical black hole,
the river has zero twist,
so objects experience no spatial rotation,
with the consequence that the metric,
the Gullstrand-Painlev\'e metric,
is flat along spatial hypersurfaces at constant time, $\dd t_\ff = 0$.
For a rotating black hole,
the river has a finite twist,
and the metric is not flat along spatial hypersurfaces.



\begin{acknowledgements}
We thank Katharina Kohler for a written translation
of the final part of Gullstrand (1922) \cite{Gullstrand},
and
Wildrose Hamilton for drawing Figure~\ref{blackholefishies}.
This paper has benefited from conversations and correspondence
with many colleagues,
including but not limited to
Peter Bender, James Bjorken, Robert Brandenberger, Jan Czerniawski,
Nick Gnedin, and Scott Pollack.
AJSH acknowledges support from NSF award ESI-0337286.
\end{acknowledgements}

\appendix
\section{Project: The river model of black holes}

The project below
has been field-tested and refined over a period of several years
in undergraduate classes on relativity and black holes
at both lower-division non-science-major
and upper-division science-major levels.
It was designed as a 45-minute ``in class group project'',
in which students would split into groups of 3 or 4,
and by arguing with each other would
arrive at consensus answers to a series of concept questions.
At the end of the project each group would submit its answers for grade.

\section*{Concept questions}

According to the river model of black holes, the behavior of
objects near black holes is precisely as if space were falling
like a river into the black hole.
For spherical black holes, this
model was discovered in 1921
by the German Nobel prizewinner Allvar Gullstrand
and independently by the French mathematician Painlev\'e.
In the model, space falls inward at the Newtonian escape velocity
$v = \sqrt{2 G M / r}$.
The infall velocity is less than the speed of light $c$ outside the horizon,
equals the speed of light $c$ at the horizon,
and exceeds the speed of light $c$ inside the horizon.

What does the river model predict for the answers to the
questions below?
[For freshman non-science majors,
use only the unstarred questions.
For more advanced, science-major students,
use all questions, and drop or abbreviate the hints.]

\begin{enumerate}

\item[$^\ast$1.]
What radius does the river model predict for the horizon of a black hole?

\item[2.]
Suppose that you are a light beam (therefore moving at the speed of light)
exactly at the horizon.
What would happen to you if were pointed directly outward?
[Do you fall in? Do you move out? Do you move sideways?]
What would happen to you if you were pointed mostly but not exactly outward?

\item[3.]
In what way, if any, does this behavior differ from
the predictions of the corpuscular theory of light,
which in the hands of John Michell in 1784
gave the ``correct'' result for the radius of the horizon?
[In the corpuscular theory of light,
a corpuscle of light is emitted at the speed of light,
and thereafter behaves much like a massive particle:
it flies outward,
and it either goes to infinity
or turns around and comes back
depending on whether its initial velocity,
the speed of light,
is more or less than the escape velocity.]


\item[4.]
Suppose that you are a light beam orbiting the black hole in a circular
orbit.  On this orbit, the so-called ``photon sphere",
are you at the horizon, inside the horizon, or outside the horizon?
Justify your answer.

\item[5.]
Make a connection between the appearance of the sky if you hover
just above the horizon of a black hole, and special relativistic beaming.
[How does a scene appear if you move through it at very close to the
speed of light?]

\item[6.]
Qualitatively, what would the river model predict
for the tidal forces experienced by an infalling observer?
[First, the tidal force in the vertical direction.
Think about the fact that the river is accelerating inward.
Next, the tidal force in the horizontal direction.
Think about the fact that the river is converging (getting narrower)
as it flows inward.]

\item[$^\ast$7.]
How does the river model account for redshifting and freezing at the horizon?

\item[$^\ast$8.]
Given that one of the fundamental propositions of Special and General
Relativity is that spacetime has no absolute existence, what does it
mean to say that space is falling into a black hole?

\item[$^\ast$9.]
In the river model, the flow of space accelerates inward to the black hole.
If the river were moving uniformly instead of accelerating,
would there be any gravity?

\end{enumerate}

%

\section*{Answers}

\begin{enumerate}
\item
The river velocity equals the speed of light when
$\left( {2 G M / r_s} \right)^{1/2} = c$,
which rearranges to an expression for the radius of the event horizon,
the Schwarzschild radius $r_s$,
\[
  r_s
  =
  {2 G M \over c^2}
  \ .
\]

\item
If you were a light beam pointed directly outward at the horizon,
then you would hang forever at the horizon,
your outward motion at the speed of light being exactly
canceled by the inward motion of the river of space at the speed of light.
If you were a light beam not exactly pointed outward,
then the outward component of your velocity would be a bit less than
the speed of light, since part of your velocity would be sideways.
The inflow of space would then carry you into the black hole.

\item
Whereas in general relativity
an outwardly pointed light beam at the horizon hangs there motionless for ever,
in the classical corpuscular theory the light never remains at rest.
The light either keeps going outward for ever
(if its velocity exceeds the escape velocity),
or else it turns around and comes back.
It is true that the light is motionless at the instant of turnaround,
but otherwise the light is always moving.

Another difference is that
in general relativity the question of
whether a light beam can escape from a point just above the horizon
depends on the direction in which the light beam is pointed.
If the light beam is pointed directly outward,
then it will escape, but if it is pointed somewhat sideways,
then it will fall into the black hole.
In the classical corpuscular theory,
by contrast, whether a corpuscle escapes from a given point
depends only on whether its velocity exceeds the escape velocity,
not on the direction in which it is pointed.


\item
You cannot be at the horizon,
because if you had any sideways motion,
which you must because you are in circular orbit,
then the inflow of space would drag you into the black hole.
And you cannot be inside the horizon,
because the inflow of space would again drag you inwards.
Therefore you must be in circular orbit somewhere outside the horizon.
For a Schwarzschild black hole,
the radius of the photon sphere turns out to be 1.5 Schwarzschild radii.

\item
If you move through a scene at very close to the speed of light,
then the scene ahead of you, in the direction you are moving,
appears concentrated, brightened, and blueshifted.
If you hover just above the horizon of a black hole,
then according to the river model you must be moving very rapidly
through the inflowing river of space.
Consequently the view above you must appear concentrated,
brightened, and blueshifted.
It should be emphasized that hovering just above the horizon
of a black hole is an unnatural and wasteful thing to do.
In reality, you would surely ``go with the flow'' of space.
If you free-fall into a black hole,
then you do not see the sky highly concentrated above you.

\item
Since the river is accelerating inwards,
the velocity of the river is faster at your feet than at your head
(presuming that you are upright,
so that your feet are closer to the black hole than your head).
The difference in river velocity
means that you feel a tidal force in the vertical direction,
pulling your feet away from your head.

In the horizontal direction,
the river is converging spherically towards the black hole,
so you feel tidally squashed in the horizontal direction.

\item
Just above the horizon, a photon battling against the inrushing
torrent of space takes a long time to get to an outside observer.
As the emitter gets closer to the horizon, it takes longer and
longer for the photon to get out,
until at the horizon it takes
an infinite time for a photon to lift off the horizon.
Thus as an object approaches the horizon,
it appears to an outside observer slower and slower,
thus more and more redshifted.
Asymptotically,
the object appears to freeze on the horizon,
and the redshift tends to infinity.

\item
The river model consists of a set of coordinates
(the background)
and a set of locally inertial frames that flow through those coordinates
(the river that flows through the background).
Attaching a set of coordinates and a set of locally inertial frames
does not make the spacetime absolute.

\item
According to the Principle of Equivalence,
a gravitating frame is equivalent to an accelerating frame,
so if there is no acceleration, then there is no gravity.
However,
if the river is falling at constant velocity in the vertical direction
but still converging horizontally because of the spherical convergence
of the flow,
then you will feel a tidal squashing in the horizontal direction,
so there must be a gravity.
\end{enumerate}

\end{document}